\documentclass[preprint]{aastex}
\usepackage{psfig}

\newcommand{\chandra}{{\it CHANDRA}}
\newcommand{\rxte}{{\it RXTE}}
\newcommand{\asca}{{\it ASCA}}
\newcommand{\rosat}{{\it ROSAT}}
\newcommand{\einstein}{{\it EINSTEIN}}
\newcommand{\ginga}{{\it GINGA}}
\newcommand{\bbxrt}{{\it BBXRT}}

\newcommand{\sax}{{\it BeppoSAX}}
\newcommand{\ec}{$\eta$ Carinae}

\slugcomment{To be Submitted to the ApJ}

\shorttitle{The X-ray Grating Spectrum of \ec}
\shortauthors{Corcoran et al.}

\begin{document}

\title{The \chandra\ HETGS X-ray Grating Spectrum of \ec}

\author{M. F. Corcoran\altaffilmark{1,2}, J. H. Swank\altaffilmark{2}, 
R. Petre\altaffilmark{2}, K. 
Ishibashi\altaffilmark{3}, K. Davidson\altaffilmark{4}, 
L. Townsley\altaffilmark{5}, 
R. Smith\altaffilmark{6}, S. White\altaffilmark{7}, R. 
Viotti\altaffilmark{8}, 
A. Damineli\altaffilmark{9}}

\altaffiltext{1}{Universities Space Research Association, 7501 Forbes 
Blvd, Ste 206, Seabrook, MD 20706} 
\altaffiltext{2}{Laboratory for High Energy Astrophysics, Goddard Space Flight 
Center, Greenbelt MD 20771} 
\altaffiltext{3}{National Research Council, Laboratory For Astronomy and Space Physics, Goddard Space 
Flight Center, Greenbelt, MD 20771} 
\altaffiltext{4}{Astronomy Department, University of Minnesota,
    Minneapolis, MN, 55455}
\altaffiltext{5}{Department of Astronomy, Pennsylvania State 
University, State College, PA}
\altaffiltext{6}{Harvard-Smithsonian Center for Astrophysics, 
Cambridge, MA}
\altaffiltext{7}{Department of Astronomy, University of Maryland,
College Park, MD, 20742}
\altaffiltext{8}{Istituto di Astrofisica Spaziale, CNR, Area di Ricerca Tor Vergata, 
00133 Roma, Italy}
\altaffiltext{9}{Instituto Astron\^{o}mico e Geofisico da USP,Av. 
Miguel Stefano 4200, 04301-904 S\~{a}o Paulo, Brazil}

\begin{abstract}
\ec\ may be the most massive and luminous star in the Galaxy and is
suspected to be a massive, colliding wind binary system.  The
\chandra\ X-ray observatory has obtained a calibrated, high-resolution
X-ray spectrum of the star uncontaminated by the nearby extended soft
X-ray emisssion.  Our 89 ksec \chandra\ observation with the High
Energy Transmission Grating Spectrometer (HETGS) shows that the hot
gas near the star is non-isothermal.  The temperature distribution may
represent the emission on either side of the colliding wind bow shock,
effectively ``resolving'' the shock.  If so, the pre-shock wind
velocities are $\sim 700$ and $\sim 1800$ km s$^{-1}$ in our analysis,
and these velocities may be interpreted as the terminal velocities of
the winds from \ec\ and from the hidden companion star.  The
forbidden-to-intercombination ($f/i$) line ratios for the He-like ions
of S, Si and Fe are large, indicating that the line forming region
lies far from the stellar photosphere.  The iron fluorescent line at
1.93\AA, first detected by \asca, is clearly resolved from the thermal
iron line in the \chandra\ grating spectrum.  The Fe fluorescent line
is weaker in our \chandra\ observation than in any of the \asca\
spectra.  The \chandra\ observation also provides an uninterrupted
high-time resolution lightcurve of the stellar X-ray emission from
\ec\ and suggests that there was no significant, coherent variability
during the \chandra\ observation.  The \ec\ \chandra\ grating spectrum
is unlike recently published X-ray grating spectra of single massive
stars in significant ways and is generally consistent with colliding
wind emission in a massive binary.

\end{abstract}

\keywords{\ec, stars: individual (\ec), stars: early-type, X-rays:stars, 
binaries: general }

\section{Introduction}

The superluminous star \ec~ has been observed by nearly every X-ray
satellite flown.  \einstein~ observations \citep{sew79,sc82,chleb84}
first resolved the star's X-ray emission from the rest of the Carina
Nebula and mapped out point-like emission centered on the star and an
elliptical X-ray ring around the star extending out to $\sim 15''$. 
\ginga\ \citep{koy90} and \bbxrt~ provided the first clear measures of
the strong Fe K line indicative of thermal emission probably produced
by shocked gas.  \rosat~ discovered the variable nature of the hard
source \citep{corc95}, while \asca~ discovered $\ge 50\times$ solar
abundance of nitrogen \citep{tsu97} in the outer homunculus and
fluorescent Fe K emission \citep{corc98} unresolved from the strong
thermal line.  A hard X-ray tail extending to $\sim 50$ keV was
observed with the \sax\ Phoswich Detector System \citep{viotti2001}.  A 2-10 keV
X-ray lightcurve obtained by \rxte\ (consisting typically of 1
observation per week since 1996) revealed small amplitude periodic
flares with $P= 85$ days \citep{corc97} and confirmed that the
variability seen in 1992 by \rosat~ recurred \citep{bish99, corc01} on
the same 5.5 year period which fits the He I 10830\AA~ line
variability reported by \citet{dam96}.  These recent X-ray
measurements along with variations in the radio \citep{dunc95} and
near-IR \citep{dam2000} all suggest that \ec\ may be the Galaxy's most
massive binary system \citep{dcl,corc01} in which the hard X-ray
emission is produced by shocked gas in the region where the wind from
\ec\ collides with the wind from the companion star.  An early
observation of \ec\ by the Advanced CCD Imaging Spectrometer (ACIS) on
\chandra\ in September 1999 \citep{sew01} provided the first image of
the X-ray regions at a resolution of $\approx 1''$, though, due to
degradation produced by charged particle damage of the CCDs, this
observation could not be precisely calibrated spectrally.

Here we report the first calibrated observation of \ec\ with the High
Energy Transmission Grating Spectrometer (HETGS, Canizares et al. 
2001, in preparation) on the \chandra\ X-ray observatory
\citep{chandra} using the ACIS linear array (ACIS-S) to read out the
dispersed spectrum.  The \chandra\ HETGS is well suited for observing
the point-like hard emission from \ec\ for two reasons: 1) the
spectral energy distribution of the emission fits nicely in the HETGS
bandpass and 2) this emission is thought to be dominated by thermal
processes which should produce strong line emission in the dispersed
spectrum.  Our deep (24.9 hour) observation of \ec\ with the HETGS
provides the first spectrally-resolved measure of the X-ray spectrum,
allowing for the first time a detailed definition of the temperature
distribution, density, and chemical abundance of the hot, unresolved
emission from the star.  In this first report we discuss the overall
spectral morphology of the unresolved X-ray emission, examine the
emission line temperature and density diagnostics, and provide an
uninterrupted lightcurve of the unresolved X-ray source over the
length of the observation.

\section{The Observation}

The \chandra\ HETGS+ACIS-S observation of \ec\ was performed on 19
November 2000 -- 20 November 2000.  The total exposure time of the
observation was 89,546 seconds.  The spacecraft roll was chosen so as
to avoid contamination of the dispersed spectra by other bright X-ray
sources in the field.  The data were cleaned and processed using the
standard pipeline processing available at the \chandra\ X-ray Center. 
Images and spectra were extracted from the level 1.5 events using the
Chandra Interactive Analysis of Observations (CIAO) analysis package.

The zero$^{th}$-order image, color-coded by X-ray energy is shown in
figure 1.  This image is similar to that previously published
\citep{sew01} though the energy calibration is better in the new
image.  In particular, the ACIS-S zero$^{th}$-order image shows the
soft elliptical ``shell'' of emission surrounding the hard X-ray core,
which is unresolved to ACIS at scales of $\sim 0.5'' \approx
10^{16}\mbox{cm} \approx 1000$ AU. The X-ray flux in the elliptical
``shell'' is very non-uniform, bright in the south and west (near the
``S-ridge'' and the ``W-condensations'' in the outer debris field,
Walborn, Blanco \& Thackary 1978) and faint in the north and east. 
The hard, unresolved ``core'' appears to be surrounded by a halo of
X-ray emission at moderate X-ray energies (though the apparent 
emission at distances between $2.5''$ and $5''$ is probably consistent 
with the point-spread of the mirror+detector, Seward et al. 2001).
 
\section{The X-ray Grating Spectrum of \ec}

The ACIS-S image of the dispersed spectrum shows that the hard source
is unresolved to \chandra; in particular, there is no observed
dispersed spectrum from any other source of emission in the ACIS-S
field aside from the unresolved, hard core source associated with \ec. 
We extracted the dispersed medium energy grating (MEG) and high energy
grating (HEG) spectra from the X-ray event file using CIAO. Both plus
and minus orders for the first, second and third order MEG and HEG
source spectra were extracted, along with appropriate background
spectra.  Figure 2 shows the MEG +1 order spectrum, while figure 3
shows the HEG +1 order in the vicinity of the Fe K line.  The MEG and
HEG spectra of \ec\ represent the first calibrated high resolution
X-ray spectra of this star uncontaminated by extended soft emission. 
The MEG spectrum shows very little emission from the unresolved source
at energies less than 1.5 keV ($\lambda > 8.5$ \AA) due to strong
absorption, and reveals a significant X-ray continuum in the $1-8$\AA\
range and the presence of strong line emission from lines of S XV-XVI,
Si XIII-XIV, Mg XII, Ca XIX, and Fe K. Strong forbidden lines of S XV,
Si XIII and Mg XII were detected.  In the HEG spectrum, the Fe line
region shows a thermal emission line produced by Fe XXV and a blend of
fluorescent Fe I K$\alpha_{1}+$K$\alpha_{2}$ lines centered at
1.94\AA.

\subsection{Modelling the Spectrum}

We attempted to fit the MEG spectrum with a combination of
optically-thin thermal emission models using the XSPEC analysis
package \citep{xspec}.  Because the dispersed spectrum has many bins
with few counts and because background does not contribute
significantly in the energy range of interest ($E > 1.5$ keV or
$\lambda < 8$ \AA), we used a modified version of the ``C-statistic''
\citep{cash79} on the total (non-background-subtracted) spectrum
instead of the $\chi^{2}$ statistic.  Previous analyses indicated that
the emission at energies above 1.5 keV could be fit by emission at a
single temperature (for eg., Corcoran et al.  2000).  As shown in
Figures 4 and 5, a variable abundance single temperature thermal model
\citep{mekal} provided a good fit to the continuum emission and most
of the resolved emission lines.  The parameters of the best fit single
component model are given in Table 1.  The single component
temperature is $kT=4.4$ keV with a column density of
$N_{H}=4.9\times10^{22}$ cm$^{-2}$, in reasonable agreement with
earlier results.  However, we found that no isothermal model could
simultaneously match the strengths of the H-like and He-like lines of
Si in the 4.5\AA\ $< \lambda <$ 6\AA\ range.  Fitting both the He-like
and H-like ions required the addition of at least one additional
thermal component.  Our best fit 2-temperature model is given in table
2 and shown in figures 4 and 5.  This 2-temperature model adequately
matches the strength of both the He-like and H-like lines.  This is
the first time that the unresolved X-ray emission from \ec\ has been
shown to require a non-isothermal temperature distribution.  The
maximum temperature we derive, $kT \approx 8.7$ keV, is nearly a
factor of 2 larger than most other published temperatures for \ec, and
to our knowledge represents the highest temperature ever associated
with an early-type star.

In our modelling of the X-ray spectrum we held most elemental
abundances at their solar values \citep{abundance}, but allowed the
abundances of Si, S and Fe (all of which have strong lines in the
X-ray spectrum) to vary.  The derived abundances for the 1 and 2
temperature models for these 3 elements are given in table 1.  In each
case the abundances of the Si, S, and Fe were slightly non-solar: Si
was found to be slightly overabundant, S significantly
overabundant (by about 70\%), while Fe was slightly underabundant.

\subsection{The He-like Lines and the $f/i$ Ratio}

The ratio of the intensity of the forbidden component to the
intercombination component of the He-like lines (the $f/i$ ratio) is a
density diagnostic \citep{gab69}, though the ratio may be also
increased by UV photoexcitation \citep{kahn01} which can suppress the
forbidden line and enhance the intercombination line.  We fit the Si
XIII, S XV and Fe XXV lines in XSPEC by first isolating the wavelength
region around the line complex of interest and using gaussians to
model the lines with inclusion of a power law component to describe
the background.  The fits are shown in figures 6--8 and the fit
parameters are given in Table 2.  The intercombination lines for all
three ions are weak and our analysis only yields upper limits for the
intensities of these lines.  We measure $f/i$ values of $>1.0$ for Si
XIII, $>2.0$ for S XV and $>2.1$ for Fe XXV. These values correspond
to electron densities of $< 10^{14}$ cm$^{-3}$ for Si XIII and $<
10^{15}$ cm$^{-3}$ for S XV and Fe XXV.

\subsection{The Fe Fluorescent Line}

We fit the Fe I fluorescent line from the HEG spectrum in XSPEC by
isolating the Fe K region and using a single gaussian line plus
power-law background component.  The line parameters are given in
Table 2.  The measured equivalent width of the line is only 39 eV, 
which is about half the value of the smallest equivalent width 
seen by  \asca\ \citep{corc00}.

\section{The X-ray Lightcurve}

Though \ec\ is known to undergo significant increases (or ``flares'')
in its 2-10 keV X-ray flux, sometimes increasing in flux by $20-50\%$
on a timescale of many days \citep{corc97,bish99}, no short-term
variability on timescales less than a day has ever been detected.  The
best previous measure of the X-ray lightcurve from \ec\ was a long
($\sim 100$ ksec) \asca\ observation spanning some 2.5 days which did
not find any significant variability in the 2-10 keV X-ray emission
\citep{corc98}.  We can re-examine the issue of \ec's short-term X-ray
variability using our \chandra\ data, since this observation provides
a unique uninterrupted view of the source for a period of about a day. 
We first extracted a lightcurve from the zeroth-order image on the S3
chip using a circular $5''$ diameter region centered on the unresolved
source.  We also extracted a background lightcurve from a source free
circular region of diameter $17''$ centered just beyond the outer
elliptical emission region surrounding the hard unresolved source. 
The net (background-subtracted) X-ray lightcurve of the unresolved
source in the zeroth order data is shown in figure 9a.  While there is
little evidence for variability in this zeroth order lightcurve, the
counting rate for the central source in zeroth-order is large enough
($\sim 0.2$ counts s$^{-1}$) that event pileup is a problem.  From the
continuous clocking data published by \citet{sew01} the unpiled rate
of the central source is 1.6 ACIS-I counts s$^{-1}$, which implies
that the unpiled rate in the ACIS-S zeroth order image is $\sim 0.8$
counts s$^{-1}$, suggesting a pileup fraction of $\sim 75\%$.  In
addition to reducing the observed count rate, this degree of pileup
will severely dampen any real source variability in the zeroth-order
data.  We used the count rate/pileup estimator tool provided by the
\chandra\ X-ray center to estimate the sensitivity to source
variability in the zeroth-order data and found that, for a source
spectrum described by the fit to the dispersed spectrum of the central
source, variations of the central source flux by about a factor of 3
imply a change in observed (piled-up) count rate of only 30\%.  Thus,
slight variations in the observed ACIS-S+HETG zeroth order lightcurve
might imply much larger flux variations in the source, if this
analysis is reliable.

To further investigate any possible variations in the hard X-ray
emission from \ec\ for the duration of the \chandra\ observation, we
extracted a lightcurve from the dispersed spectrum of the central
source.  Since the dispersed spectrum is spread over a large number of
detector pixels, event pileup is not significant.  To avoid problems
caused by differing chip sensitivities and backgrounds, we considered
only the MEG minus-order data from the ACIS-S2 chip.  We extracted all
dispersed counts in the MEG-minus order from a narrow rectangular
region centered on the dispersed spectrum, and extracted background in
a similar region offset from the dispersed spectrum.  In the region
used for source extraction the maximum number of counts per unbinned
ACIS-S2 pixel was 19 counts, which corresponds to an observed count
rate of $2\times 10^{-4}$ counts s$^{-1}$, or $6 \times 10^{-4}$
counts per readout frame.  Figure 9b shows the net lightcurve of the
dispersed MEG minus-order spectrum from chip S2.  This lightcurve
shows little evidence of variability, suggesting that \ec\ did not
undergo any real coherent variations during the \chandra\ observation.

\section{Discussion}

Recent evidence suggests that \ec\ may be a massive binary star, in
which variable X-ray emission is produced by shocked gas at the
interface where the wind from \ec\ collides with the wind of its
companion (presumably some less massive early-type star).  The line
complexes resolved in the HETGS spectra provide unique information
about the physical condition of the X-ray emitting region and thus on
the single or binary nature of \ec.  In the colliding wind model the
wind from \ec\ forms a bow shock around the companion, since the
companion's wind is probably weaker than \ec's wind, i.e. $\eta \equiv
(\dot{M} V_{\infty})_{c}/(\dot{M} V_{\infty})_{\eta} < 1$; here
$\dot{M}$ represents the wind mass loss rate and $V_{\infty}$ the wind
terminal velocity, and the subscripts $\eta$ and $c$ refer to \ec\ and
the companion, respectively.  In published models
\citep{dam2000,corc01} the mass loss rate from \ec\ is thought to be
at least a factor of 10 larger than that of the companion, while the
terminal velocities are probably within a factor of 3 (see below), so
that $\eta < 0.3$.

The mere presence of numerous strong emission lines confirms that the
X-ray emission is dominated by thermal processes from shocked gas. 
The dominance of thermal processes in the \chandra\ energy range
suggests that there is little or no contribution from non-thermal
processes (which could be produced via inverse compton scattering of
photospheric UV photons if Fermi acceleration of the electron
population in the shock is important).  The temperature distribution
of the X-ray emission as measured by \chandra\ provides a measure of
the pre-shock wind velocities for both stars.  In a colliding wind
binary, the temperature of the shocked gas facing either star is $kT\
\mbox{(keV)} \approx 2.59 \mu V^{2} $ (see the eq.  51 in
\cite{usov92}) where $\mu$ is the mean mass per particle and $V$ is
the stellar wind speed in 1000 km s$^{-1}$.  For simplicity we assume
$\mu \approx 1$, though this value may not be truly representative of
the chemical composition of the wind of either $\eta$ Carinae or its
companion.  We neglect the motion of the shock stagnation point due to
the orbital motion of the stars, since orbital motion has a
significant influence on the derived temperatures only when the stars
are near periastron (and the stellar velocities are greatest), while
the HETGS observation was obtained near apastron according to the
ephemeris of \citet{corc01}.  The temperatures we derived, $kT \approx
1.1$ and $\approx 8.7$ keV, suggest velocities of $\sim 700$ km
s$^{-1}$ and $\sim 1800$ km s$^{-1}$.  The speed of the wind from \ec\
is thought to be $\sim 500$ km s$^{-1}$ \citep{hill01}, fairly
consistent with the lower velocity we derive from our X-ray analysis. 
A velocity of $\sim 500$ km s$^{-1}$ is sufficient to produce the $kT
\sim 1$ keV emission, but cannot produce the hotter emission revealed
in the MEG spectrum.  Naively, we interpret the low-temperature
component as shocked emission from the relatively slow wind of \ec,
and the high-temperature component as shocked emission by the faster
wind from the companion star, so that $V_{\infty,\eta}\approx 700$ km
s$^{-1}$ and $V_{\infty,c}\approx 1800$ km s$^{-1}$.

The $f/i$ ratios derived from our analysis of the He-like ions are
large, and show that the line forming region has a density
$n_{e}<10^{14}$ cm$^{-3}$ and that enhancement of the intercombination
component by UV photoexcitation is unimportant.  This suggests that
the line forming region is located far from the stellar photosphere,
where wind densities and UV fluxes are low.  This is broadly
consistent with published descriptions of the orbit
\citep{dam2000,corc01} in which the shocked emission forms far from
either star (though closer to the companion than to \ec) in a region
of low particle and UV photon density.  In the colliding wind model,
the apex of the bow shock is located at a distance of
$d/(1+\sqrt{\eta})$ from \ec, where $d$ is the separation of the two
stars.  At the phase of the \chandra\ observation, $\phi=0.60$, the
two stars are separated by about $4\times 10^{14}$ cm, using the
orbital elements of \citet{corc01}.  At this time the bow shock is
about $2.5\times 10^{14}$ cm from \ec, and about $1.5\times 10^{14}$
cm from the companion.  The density of the wind at a distance $r$ from
the photosphere is $n = \dot{M}/(4 \pi r^{2} V\mu m_{H})$, where
$\dot{M}$ is the mass loss rate and $V$ the wind velocity.  Using
$\dot{M}_{\eta} = 10^{-4} M_{\odot}$ yr$^{-1}$ and $V_{\infty,\eta}
\approx 500$ km s$^{-1}$ \citep{hill01}, then the density of the wind
from \ec\ at a distance of $2.5\times10^{14}$ cm is only $n\approx
8\times 10^{8}$ cm$^{-3}$.  Using $\dot{M}_{c}=10^{-5} M_{\odot}$
yr$^{-1}$ \citep{corc01} and $V_{\infty,c}=1800$ km s$^{-1}$ as
appropriate values for the companion's wind, the density of the
companion's wind at a distance $r=1.5\times 10^{14}$ cm is also only
about $10^{8}$ cm$^{-3}$.  The predicted densities are consistent with
the upper limits derived from our analysis of the $f/i$ line intensity
ratios.  However, if the actual density of the emission region is near
the limit implied by the $f/i$ ratios, this might indicate that the
assumed wind momentum balance is not quite right.

In contrast, newly published X-ray grating spectra of $\theta^{1}$ Ori
C \citep{schulz00}, $\zeta$ Ori \citep{waldron01} and $\zeta$ Pup
\citep{kahn01} all have $f/i$ ratios which are lower than the values
we derive from the \ec\ X-ray line spectrum.  None of these stars are
known to show any colliding wind effects, and apparently, in these
massive stars, the X-ray lines form relatively near the stellar
photosphere, probably within a few stellar radii.  This is not the
case for \ec.

The strength of the Fe fluorescent line is related to the column
density of scattering material by $EW \approx 2.3 N_{24}$ keV, where
$EW$ is the equivalent width of the line in keV and $N_{24}$ the total
column density of cold material in units of $10^{24}$ cm$^{-2}$
\citep{kall95}.  The \asca\ spectra suggest that $EW \approx 4-7
N_{24}$ for \ec\ outside of eclipse \citep{corc00}.  Thus at the time
of the HETGS observation the equivalent width of the Fe fluorescent
line implies $N_{24}\approx 0.01-0.03$, i.e. that the column density
of cold material is $N_{H}\approx 1-3\times 10^{22}$ cm$^{-2}$.  This
value is in fair agreement with the value of $N_{H}= 5\times 10^{22}$
derived from fitting the X-ray continuum in the MEG spectrum,
suggesting that some of the same material which produces the X-ray
absorption also is responsible for producing the Fe fluorescent
emission.  The equivalent width of the line which we derive here is
much smaller than any values given in either \citet{corc00} or
\citet{corc98}.  This may be the result of the difficulty in
determining the actual width of the fluorescent line in the \asca\
spectra since the line is not resolved from the thermal component. 
However, if this variation is real, it may represent a real decrease
in the scattering optical depth.  Such a decrease is not unexpected in
the binary model since at the time of the HETGS observation, the
companion is nearly in front.  Since the companion's wind is less
dense this could produce a decrease in the scattering optical depth at
the time of the HETGS observation.

\section{Conclusions}
We have presented here the first high-energy X-ray grating spectrum of
the supermassive star \ec.  This grating spectrum shows strong line
emission from H-like and He-like ions of S, Si, Mg, Ca, and Fe, and
confirms the presence of the Fe fluorescent line discovered by \asca. 
The forbidden lines of the He-like ions are strong and the
intercombination lines weak, indicating that the line forming region
is far from the stellar photosphere.  These results are all consistent
with the current picture of \ec\ as a colliding wind binary.  We
expect that interesting variations in the X-ray spectrum will occur as
the two stars approach periastron (which should next occur on June 20,
2003).  In particular, we expect that the strength of the forbidden
lines will weaken near periastron, as the density in the X-ray region
increases and as the UV photospheric flux in the line region
intensifies.  Additional HETGS observations as the stars approach
periastron, especially in conjunction with UV spectroscopy, will
provide unique and significant constraints on the orbital geometry,
the strengths of the winds from both stars, and the evolutionary
stages of the stellar components.

\acknowledgements We would like to thank Fred Seward for important
contributions to this paper, in particular in emphasizing the
importance of pileup in measuring the variation of the central source. 
We also gratefully acknowledge the \chandra\ staff at the SAO/CfA and
the HETGS team at MIT for their help in scheduling the observation and
in analyzing the data.  We made use of the FTOOLS suite of software
supported by the HEASARC. This research has made use of NASA's
Astrophysics Data System Abstract Service.

\clearpage 

\noindent
Figure 1.  Zeroth order image, color coded for 
X-ray energy (red - low energy; green - moderate energy; blue - high 
energy). The diagonal band at the top of the image is the dispersed 
medium energy grating (MEG) spectrum.

\noindent Figure 2.  +1 order Medium Energy Grating (MEG) spectrum of
\ec.  Strong lines from H- and He-like ions of Mg, Si, S, Ca and Fe
are indicated.  For comparison the spectrum of the background is also
shown; background only contributes about 1 count in each spectral bin. 

\noindent Figure 3.  Iron K line region from the +1 order High Energy
Grating (HEG) spectrum of \ec.  Emission is seen from He-like iron
originating in gas at a temperature of $\sim$ 8 keV. In addition
fluorescent line emission from cool iron is also clearly detected. 
The spectrum at the bottom is the X-ray background spectrum.

\noindent
Figure 4. Fits to the MEG +1 order spectrum.  A
two-temperature fit using a variable abundance collisionally ionized
plasma model \citep{mekal} is shown in blue, and a single-temperature
fit with the same emission model is in red.  The single temperature
model cannot reproduce simultaneously the strength of the S XV, Si
XIII and Si XIV lines in the range 4.5--7.0\AA. 

\noindent 
Figure 5. Same as in Figure 4, but 
emphasizing the S \& Si line region. 

\noindent
Figure 6. Fit to the $f,i,r$ lines of Si 
XIII.

\noindent
Figure 7. Fit to the $f,i,r$ lines of S XV.

\noindent
Figure 8. Fit to the $f,i,r$ lines of Fe XXV (near 
1.85\AA) and the Fe I fluorescent line (at 1.93\AA).

\noindent
Figure 9a. The background-corrected lightcurve of the
unresolved hard source derived from the zeroth-order data.  Event
pileup is severe (the pileup fraction is about 75\%), which will
severely reduce the observed amplitude of any real source
variability.

\noindent 
Figure 9b. The background-corrected lightcurve of the
unresolved hard source derived from the dispersed MEG minus order for
events falling on the S2 chip.  Event pileup is not significant in the
dispersed data. 

\clearpage

\begin{figure}[tbp]
 \figurenum{Figure 1}
 \centering
 \psfig{file=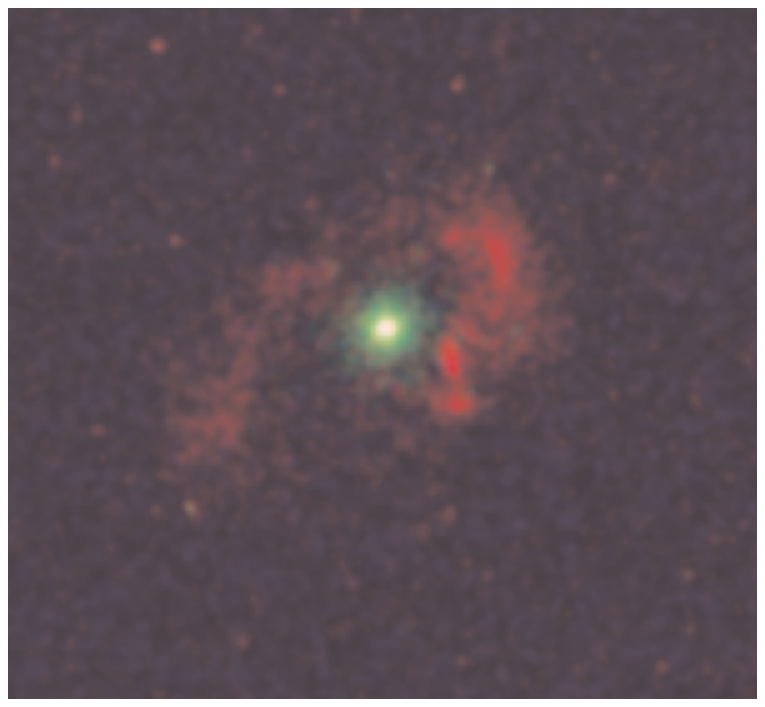,height=8in,angle=0}
 \end{figure}

 \clearpage
 
 \begin{figure}[tbp]
 \figurenum{Figure 2}
 \centering
 \psfig{file=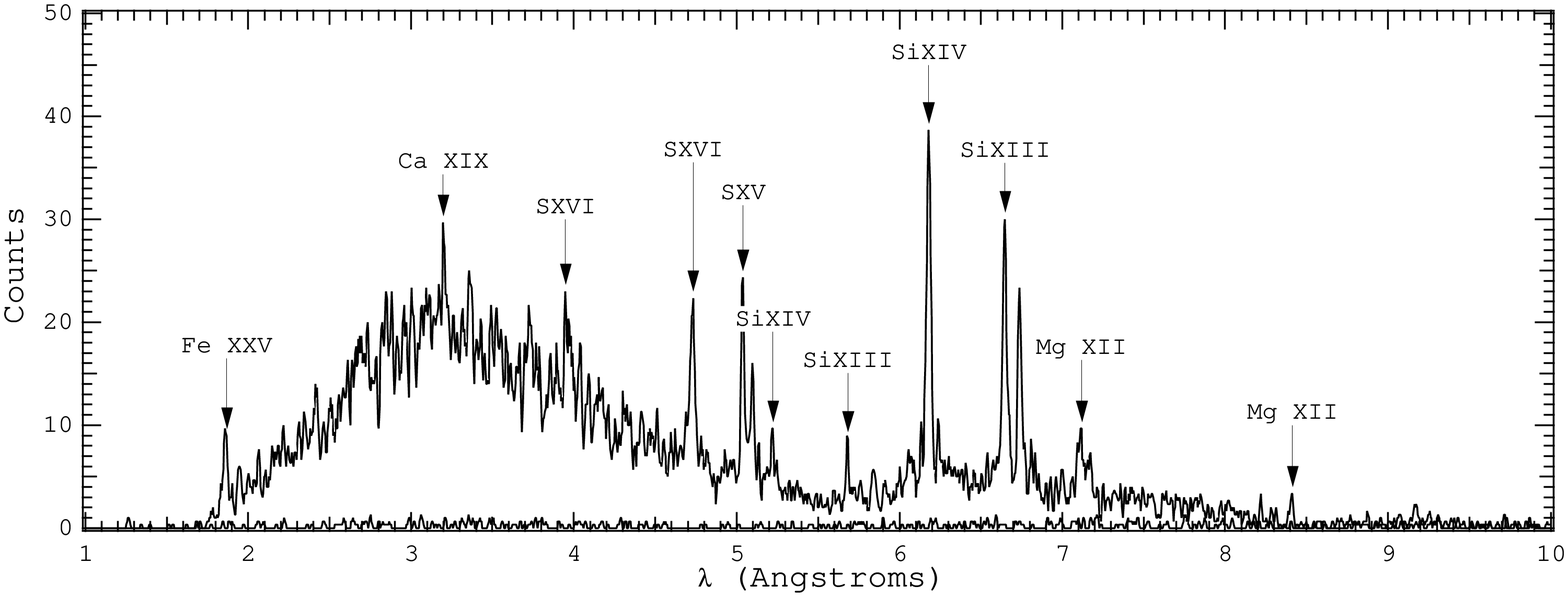,height=8in,angle=90}
 \end{figure}

 \clearpage
 
 \begin{figure}[tbp]
 \figurenum{Figure 3}
 \centering
 \psfig{file=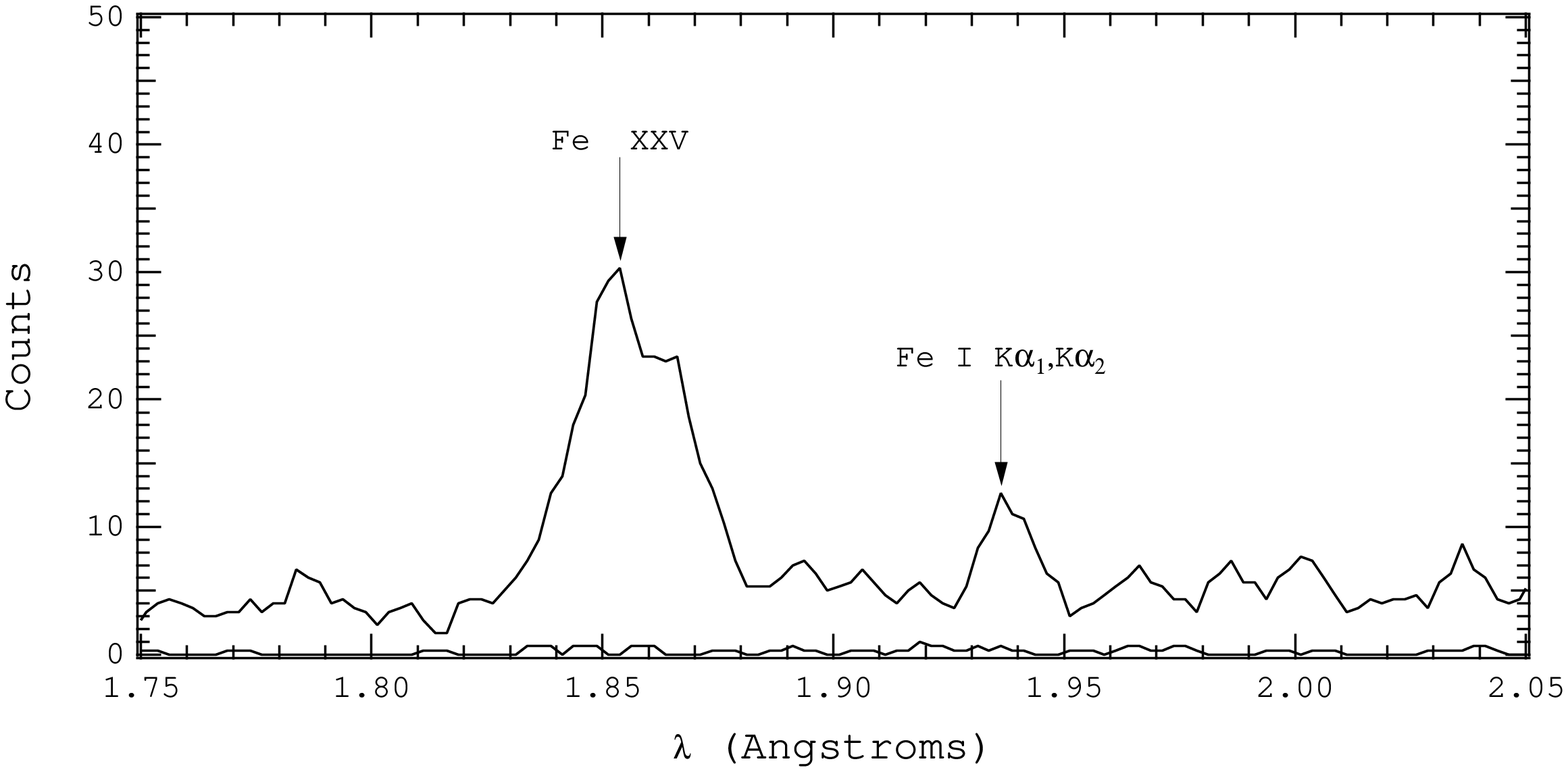,height=8in,angle=90}
 \end{figure}

  \clearpage
 
 \begin{figure}[tbp]
 \figurenum{Figure 4}
 \centering
 \psfig{file=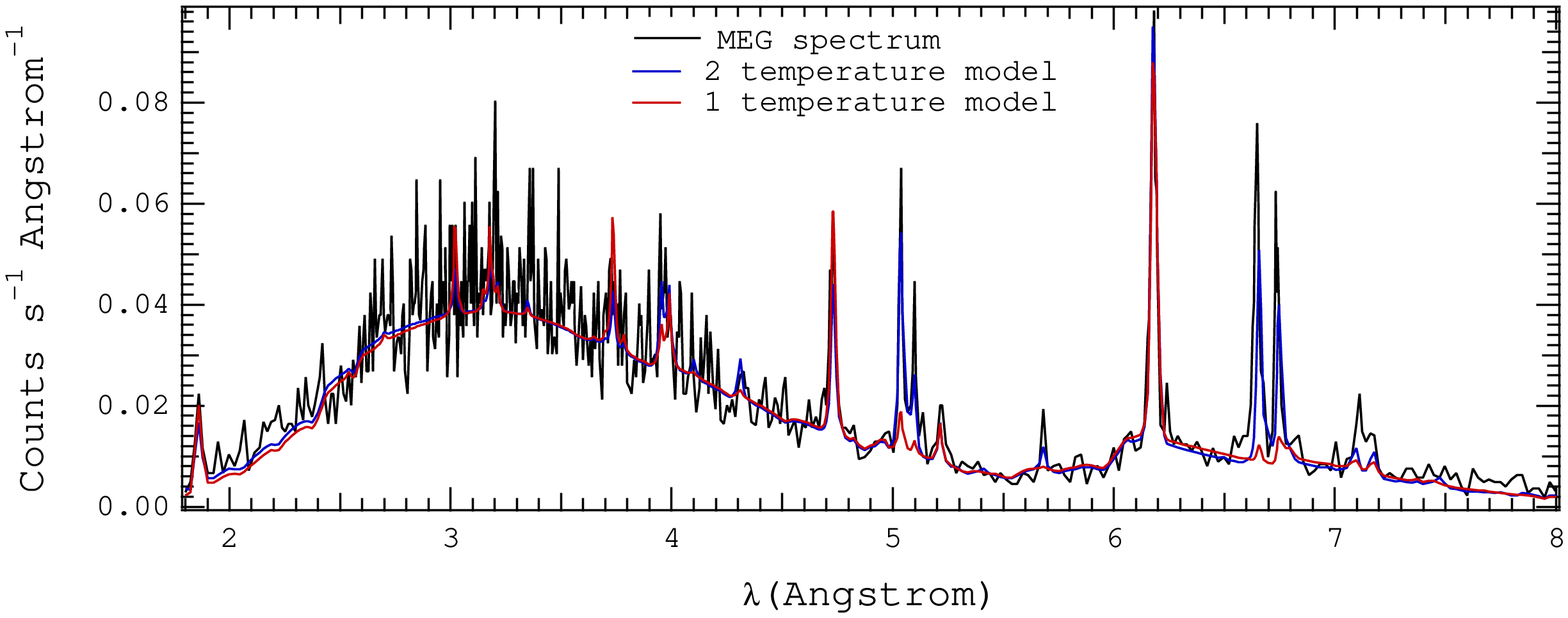,height=8in,angle=90}
 \end{figure}

   \clearpage
 
 \begin{figure}[tbp]
 \figurenum{Figure 5}
 \centering
 \psfig{file=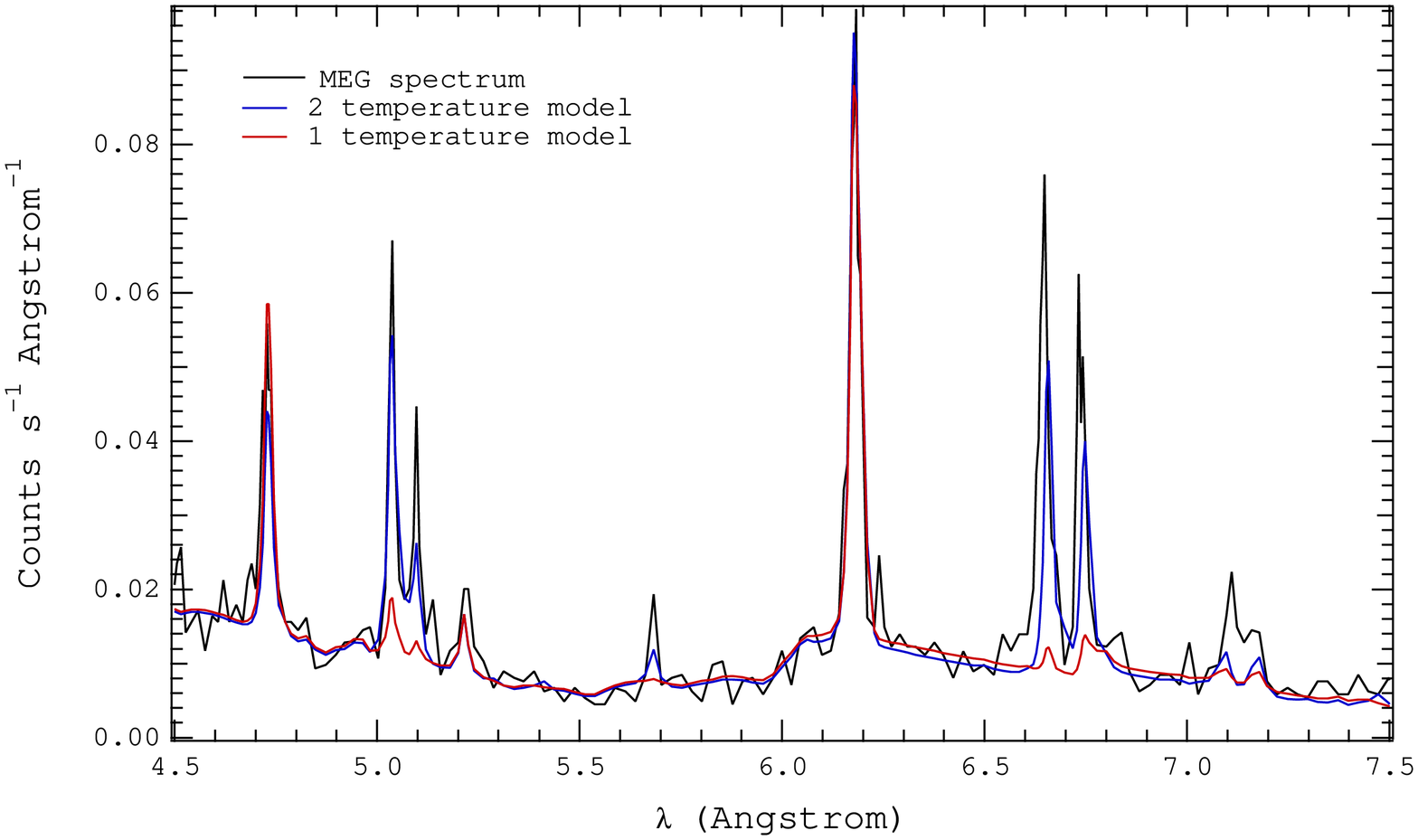,height=8in,angle=90}
 \end{figure}

    \clearpage
 
 \begin{figure}[tbp]
 \figurenum{Figure 6}
 \centering
 \psfig{file=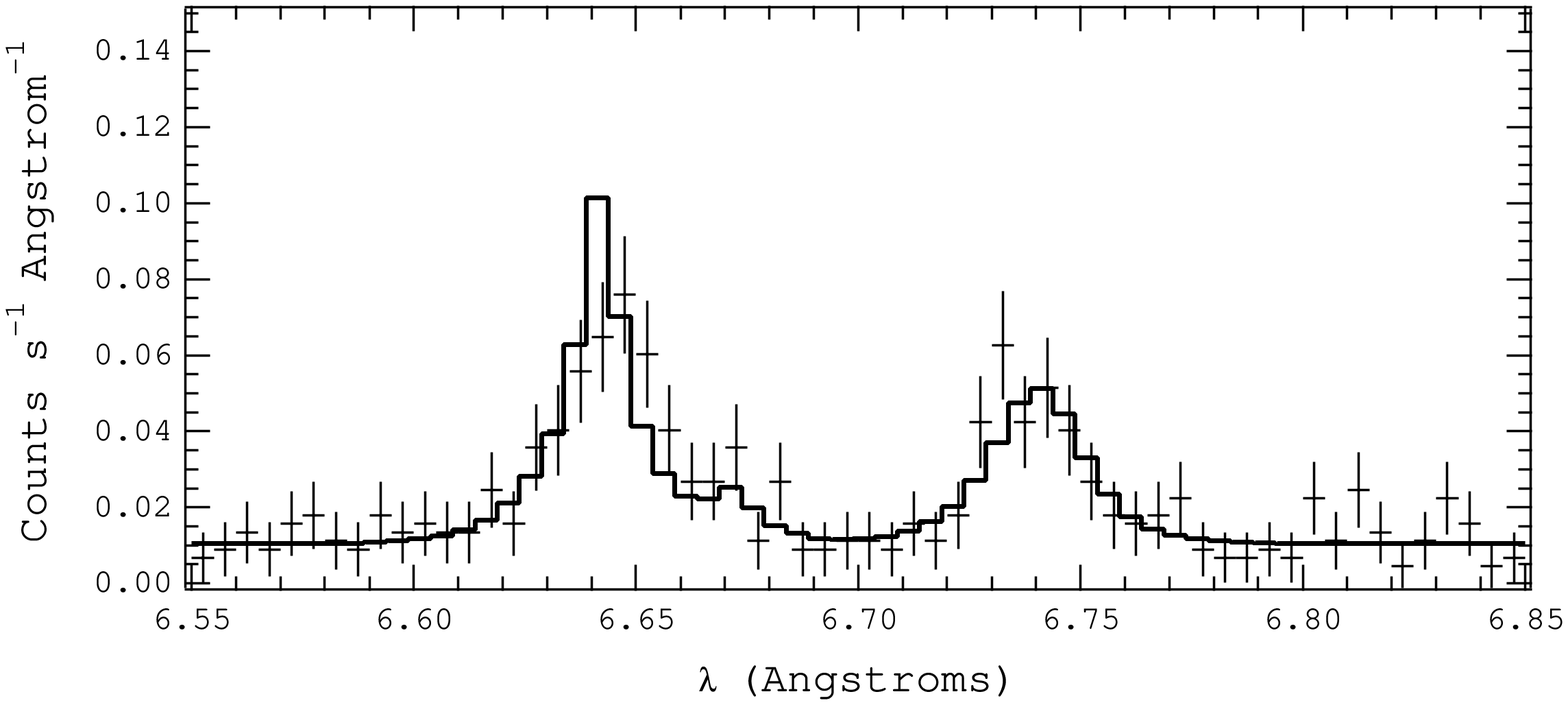,height=3in,angle=0}
 \end{figure}

    \clearpage
 
 \begin{figure}[tbp]
 \figurenum{Figure 7}
 \centering
 \psfig{file=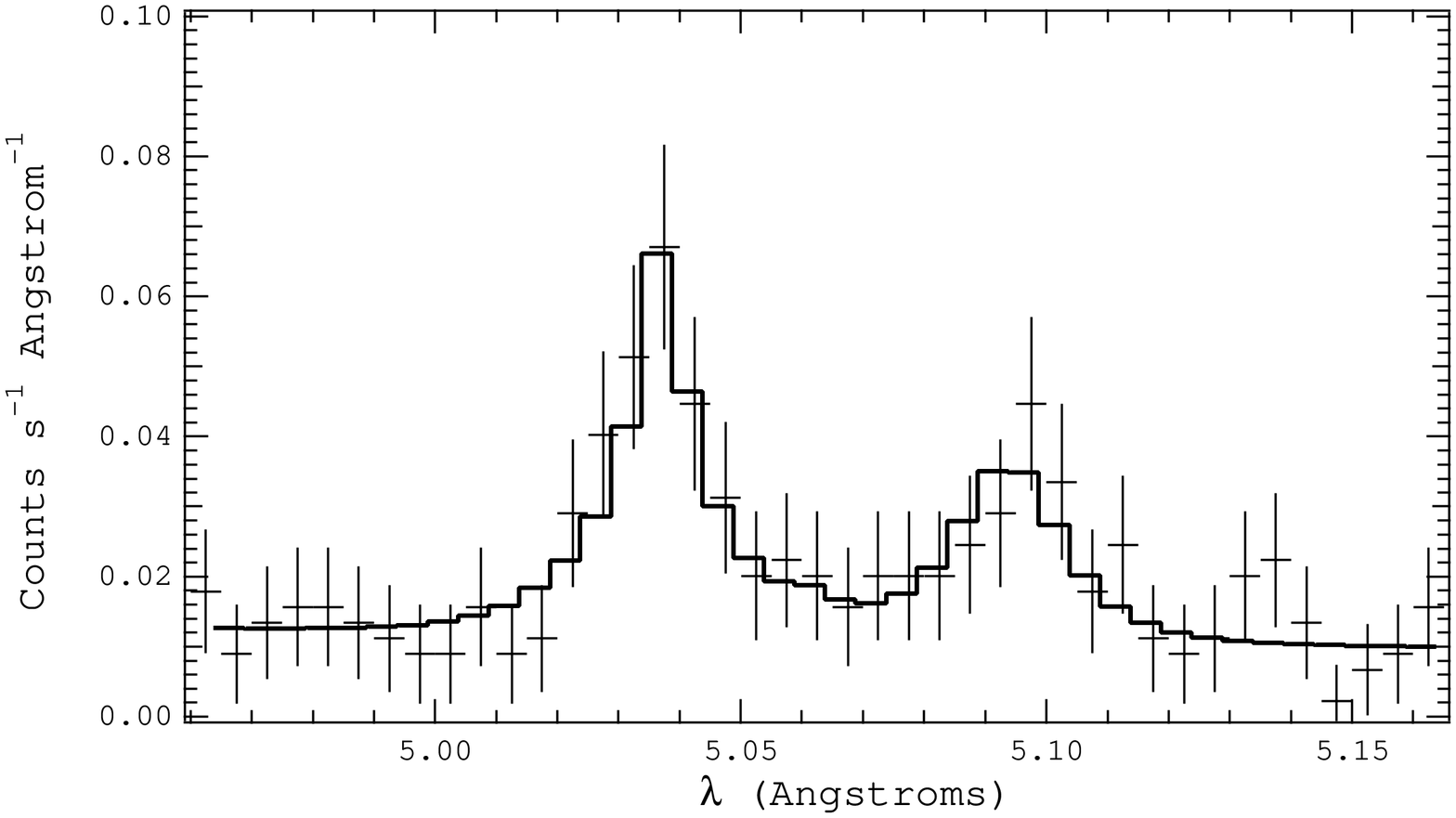,height=3in,angle=0}
 \end{figure}
 
    \clearpage
 
 \begin{figure}[tbp]
 \figurenum{Figure 8}
 \centering
 \psfig{file=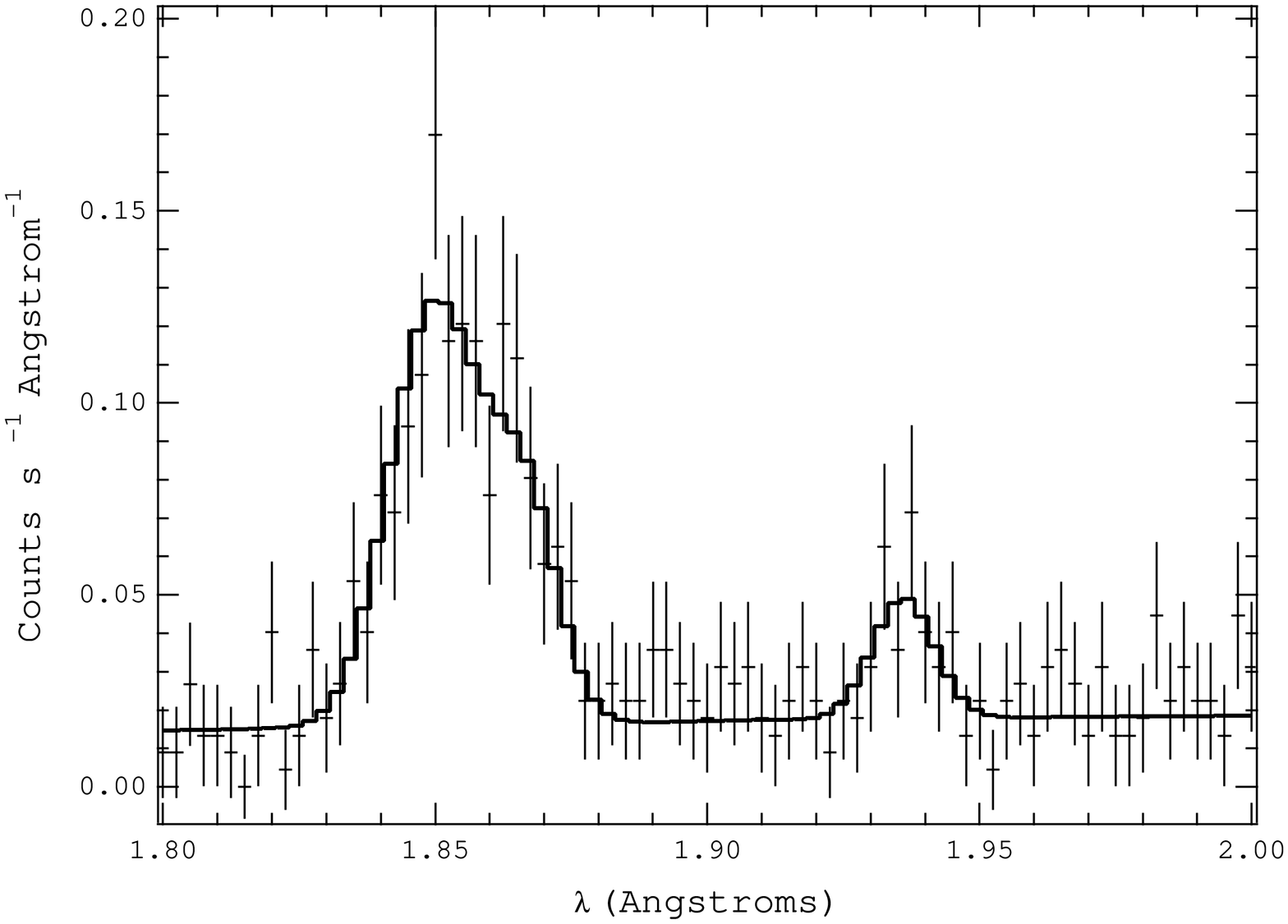,height=3in,angle=0}
 \end{figure}

 \clearpage
 
 \begin{figure}[tbp]
 \figurenum{Figure 9a}
 \centering
 \psfig{file=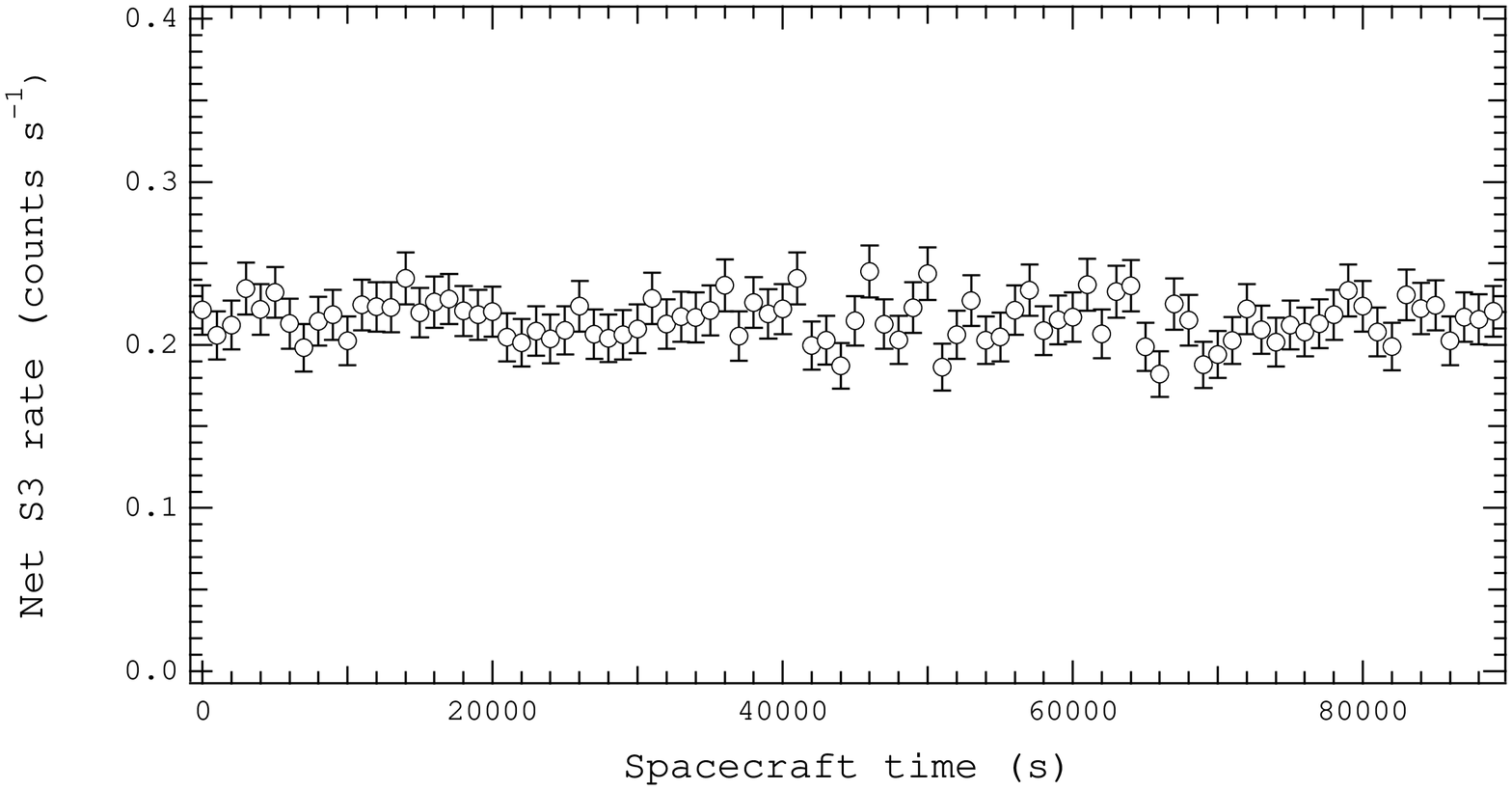,height=3in,angle=0}
 \end{figure}

 \clearpage
 \figurenum{Figure 9b}
 \begin{figure}[tbp]
 \centering
 \psfig{file=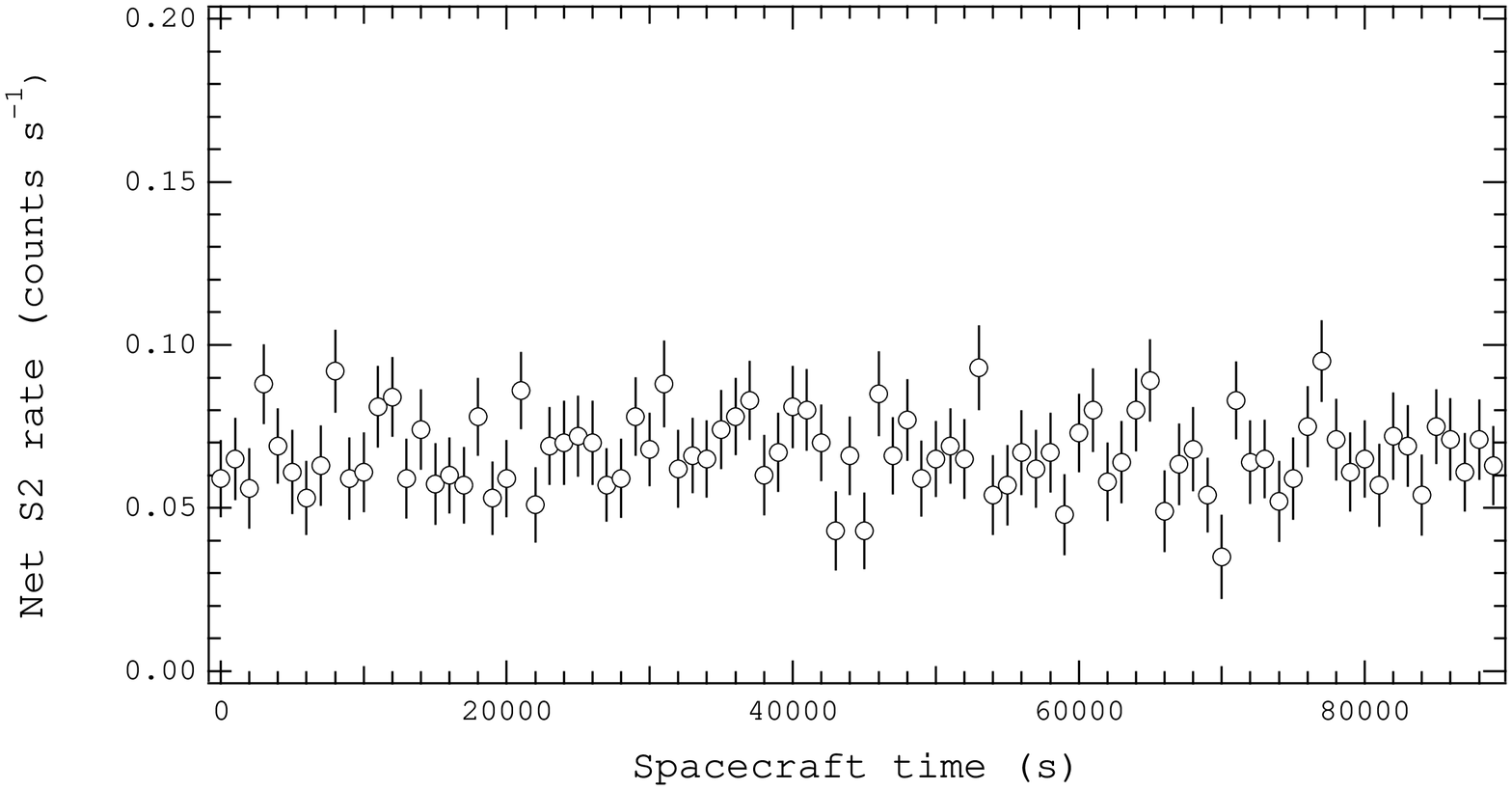,height=3in,angle=0}
 \end{figure}

 \clearpage

 \begin{center}
\begin{tabular}{lccc}
    \multicolumn{4}{c}{Table 1.} \\
    \hline
\multicolumn{1}{l}{Parameter} & \multicolumn{1}{c}{Units} & 
\multicolumn{1}{c}{1 Component Fit} & \multicolumn{1}{c}{2 Component Fit}\\
\hline
$N_{H}$                     & $10^{22}$cm$^{-2}$     & $4.9$    & $5.1(\pm0.4)$ \\
$kT$                        & keV                    & 4.4      & $1.1(\pm0.1)$ \\
Emission Measure$^{1}$      & $10^{57}$cm$^{-3}$     & 4.0      & $1.8(\pm0.1)$ \\
Lx (2-10, absorbed)$^{1}$   & ergs s$^{-1}$          & $2.6$    & $3.0$ \\
Lx (2-10, unabsorbed)$^{1}$ & $10^{34}$ergs s$^{-1}$ & $3.9$    & $4.5$ \\
Si/Si$_{\odot}$             &                        & 1.5      & $1.1(\pm0.2)$ \\
S/S$_{\odot}$               &                        & 1.4      & $1.7(\pm0.4)$ \\
Fe/Fe$_{\odot}$             &                        & 0.6      & $0.9(\pm0.2)$ \\
$kT_{2}$                    & keV                    &          & $8.7(\pm1.7)$ \\
Emission Measure$^{1}~_{2}$  & $10^{57}$cm$^{-3}$     &          & $2.9(\pm0.3)$ \\
$C-$statistic               &                        &  2035                  & 1619 \\
\hline
\multicolumn{3}{l}{$^{1}$ calculated assuming a distance of 2100 pc \citep{corc01}} \\
\end{tabular}
\end{center}

      \clearpage
    
 \begin{center}
\begin{tabular}{lcccc}
    \multicolumn{5}{c}{Table 2.} \\
    \hline
\multicolumn{1}{c}{Line} & \multicolumn{1}{c}{$E$ (keV)} &
\multicolumn{1}{c}{Measured $E$ (keV)} & \multicolumn{1}{c}{Eq. Width (eV)} &
\multicolumn{1}{c}{Intensity} \\
& & & & (photons s$^{-1}$ cm$^{-2}$) \\
\hline
SiXIII (r) & 1.864 & 1.866 & 38 & 0.009 \\
SiXIII (i) & 1.853 & $1.859$ & $<4.3$ & $<0.002$ \\
SiXIII (f) & 1.839 & 1.839 & 21.4 & 0.002 \\
SXV (r) & 2.460 & 2.460 & 33.1 & 0.009 \\
SXV (i) & 2.450 & $2.450$ &$<6.0$ & $<0.001$ \\
SXV (f) & 2.431 & 2.434 &25.6 & 0.002 \\
FeXXV (r) & 6.700 & 6.700 & 261 & 0.034 \\
FeXXV (i) & 6.682 &  $6.680$ & $<42$ & $<0.007$ \\
FeXXV (f) & 6.636 & 6.640 & 42  & 0.015 \\
FeI   & 6.424 & 6.407 & 39.4 & 0.009 \\
\hline
\end{tabular}
\end{center}

\end{document}